# A Review on Industrial Augmented Reality Systems for the Industry 4.0 Shipyard

PAULA FRAGA-LAMAS[1], (Member, IEEE), TIAGO M. FERNÁNDEZ-CARAMÉS[1], (Senior Member, IEEE), ÓSCAR BLANCO-NOVOA[1], and MIGUEL VILAR-MONTESINOS[2]
[1] Unidad Mixta de Investigación Navantia-UDC, Universidade da Coruña, Edificio de Talleres Tecnológicos, Mendizábal, s/n, 15403, Ferrol, Spain (e-mail: paula.fraga@udc.es; tiago.fernandez@udc.es; o.blanco@udc.es)
[2] Navantia S. A., Unidad de Producción de Ferrol, Taxonera, s/n, 15403, Ferrol, Spain. (e-mail:mvilar@navantia.es)

Corresponding authors: Tiago M. Fernández-Caramés and Paula Fraga-Lamas (e-mail: tiago.fernandez@udc.es; paula.fraga@udc.es).

This work is part of the Plant Information and Augmented Reality research line of the Navantia-UDC Joint Research Unit and has been funded by it.

**ABSTRACT** Shipbuilding companies are upgrading their inner workings in order to create Shipyards 4.0, where the principles of Industry 4.0 are paving the way to further digitalized and optimized processes in an integrated network. Among the different Industry 4.0 technologies, this article focuses on Augmented Reality, whose application in the industrial field has led to the concept of Industrial Augmented Reality (IAR). This article first describes the basics of IAR and then carries out a thorough analysis of the latest IAR systems for industrial and shipbuilding applications. Then, in order to build a practical IAR system for shipyard workers, the main hardware and software solutions are compared. Finally, as a conclusion after reviewing all the aspects related to IAR for shipbuilding, it is proposed an IAR system architecture that combines Cloudlets and Fog Computing, which reduce latency response and accelerate rendering tasks while offloading compute intensive tasks from the Cloud.

## I. INTRODUCTION

The shipbuilding industry is aimed at providing integral solutions for delivering fully operational vessels together with their life-cycle maintenance. Shipbuilding main working areas include, among others, the design and construction of hi-tech vessels, the development of naval control and combat systems, overhauls of military and civil vessels, ships repairs, and diesel engine and turbine manufacturing. In all those areas, most companies are transferring the Industry 4.0 principles to their shipyards in order to build Shipyards 4.0, seeking to apply the newest technologies related to ubiquitous sensing, Internet of Things (IoT), robotics, Cyber-Physical Systems (CPS), 3D printing or Big Data to improve the efficiency of the many processes that occur in a shipyard.

This paradigm shift will change the way that operators perform their daily tasks. Therefore, they must be equipped with devices that act as an interface for human-machine communication and collaboration, and even as a Decision Support System (DSS) that would help to optimize their actions. Augmented Reality (AR), and specifically IAR, is one of the technologies that provide powerful tools that support the operators that undertake tasks, helping them in assembly tasks, context-aware assistance, data visualization and interaction (acting as a Human-Machine Interface (HMI)), indoor localization, maintenance applications, quality control or material management. Specifically, AR technology is expected to grow significantly in the next years together with Virtual Reality (VR), creating a market of US $80bn in 2025 [1].

IAR is one of the technologies analyzed by Navantia, one of world's largest shipbuilders. At the end of 2015 Navantia created together with the University of A Coruña (UDC) the Joint Research Unit Navantia-UDC, which studies the application of different Industry 4.0 technologies to shipyards through various research lines. One of such lines is called "Plant Information and Augmented Reality" and has among its objectives to evaluate the feasibility of using IAR to provide information to operators about shipyard processes.

The present paper is aimed at analyzing the latest research and the best technologies to build an IAR system for a shipyard. Specifically, its main contributions are the following, which, as of writing, have not been found together in the literature:

- The article reviews the main characteristics of the most recent IAR systems for industrial and shipbuilding applications.
- It provides thorough comparisons on the latest hardware and software technologies for creating IAR solutions.
- This article also identifies and discusses different use cases where IAR can be useful for shipbuilding.
- It proposes a novel IAR architecture that makes use of Edge Computing to reduce latency response and accelerate rendering tasks.

The remainder of this paper is organized as follows. Section II reviews the main IAR concepts to provide a basic understanding of the underlying technology. Section III summarizes the main industrial applications of IAR, while Section IV analyzes the main academic and commercial IAR developments for the shipbuilding industry. Section V describes the most relevant use cases of IAR for a shipyard. Section VI compares the main IAR hardware and software technologies for developing shipbuilding applications. Section VII analyzes the traditional IAR architecture and then proposes the ideal architecture for a Shipyard 4.0. Finally, Section VIII is devoted to the conclusions.

## II. IAR BASICS
### A. ORIGINS, CURRENT TRENDS AND PROSPECTS OF IAR

The term IAR can be defined as the application of AR in order to support an industrial process [2]. This definition is broader than the one given by other researchers [3], who disregarded photo-based augmentations, which have proved to be effective in industrial environments.

Although the origins of AR can be traced back to the 1960s, when Ivan Sutherland performed his visionary experiments [4], [5], the concept of IAR actually emerged in the early 1990s, when Caudell and Mizell [6], from Boeing, presented a head-up see-through display used to augment the visual field of an operator with information related to the task she/he was carrying out. It was not until the end of the 1990s when IAR started to gain some traction, mainly thanks to the German government, who funded the ARVIKA project [7], [8]. In such a project companies like Airbus, EADS, BMW or Ford participated in the development of mobile IAR applications [9]. Such a project represented an important milestone and promoted diverse IAR initiatives. The research efforts of ARVIKA derived in 2006 in the ARTESAS (Advanced Augmented Reality Technologies for Industrial Service Applications) project [10], which focused on the development of IAR technologies for automotive and aerospace maintenance.

The first IAR systems were mostly experimental, but in the last years they started to move away from research and several relevant commercial initiatives have been launched. For instance, in 2013 Google Glasses [11] caught the attention of a broad audience that was not familiarized with AR or IAR and gained the enthusiasm of certain industries. Since then, technological giants have been actively developing AR prototypes and products [12]. Moreover, ABI Research [13] anticipates a fast-growing market with shipments forecast to expand from just over one million units in 2016 to more than 460 million by 2021 [14]. In addition, according to IDC [15], worldwide spending on AR and VR is forecast to reach US$ 17.8 billion in 2018 [16]. Furthermore, products and services will continue to grow at a similar rate throughout the remainder of the 2017-2021 forecast period, achieving a five-year Compound Annual Growth Rate (CAGR) of 98.8%. Nevertheless, according to numerous sources [17]–[19], it will take a few years until IAR reaches the maturity level to be fully deployed in industrial applications.

### B. ESSENTIAL PARTS OF AN AR SYSTEM

AR involves a set of technologies that make use of an electronic device to view, directly or indirectly, a real-world physical environment that is combined with virtual elements. The elements that make up an AR system are:

- An element to capture images (i.e., a Charge-Coupled Device (CCD), stereo, or depth-sensing camera).
- A display to project the virtual information on the images acquired by the capture element. Display technologies are divided basically into two types: video-mixed (example in Figure 1) and optical see-through displays [20] (e.g., projection-based systems). There is also a growing interest in retinal projection, but its use is currently very rare in industrial environments. In a video-mixed display, the virtual and real information previously acquired with a camera are digitally merged and represented on a display. This technology presents, among others, the disadvantages of providing limited fields of vision and reduced resolutions. In contrast, in an optical see-through display, virtual information is superimposed on the user's field of view by means of an optical projection system. The hardware used by these two display technologies can be classified as:

- Hand-Held Displays (HHD). They embed a screen that fits in a user's hand (e.g., tablets, smartphones).
- Spatial Displays. They use digital projectors to show graphic information about physical objects. These displays ease collaborative tasks among users, since they are not associated with a single user.
- Head-Mounted Displays (HMD). They are the displays included in devices like smart glasses and smart helmets, which allow users to see the entire environment that surrounds them.
- A processing unit that outputs the virtual information to be projected.
- Activating elements (e.g., images, GPS positions, QR markers, or sensor values from accelerometers, gyroscopes, compasses, altimeters or thermal sensors) that trigger the display of virtual information.

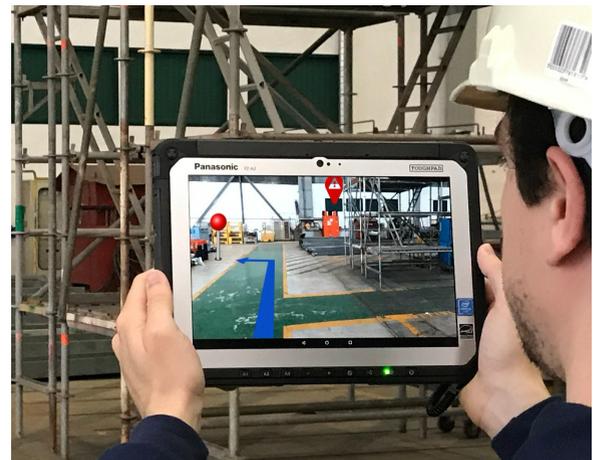

FIGURE 1: Example of video-mixed display.

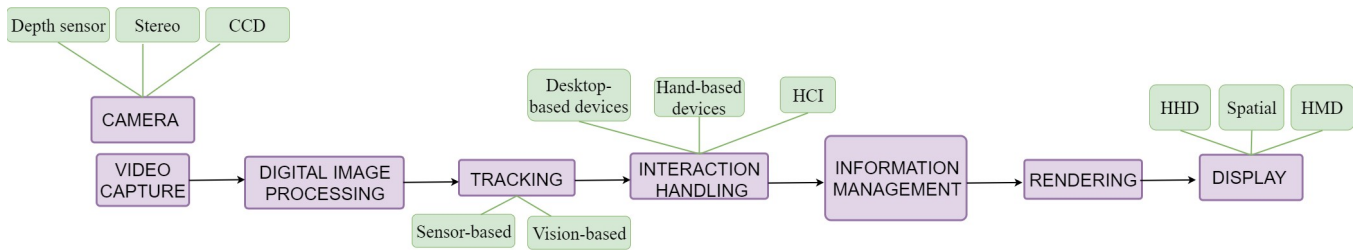

FIGURE 2: Simplified AR pipeline.

Regarding the internal logic of an AR system, it is essentially composed by the functional modules shown in Figure 2. In such a Figure it is represented a pipeline where, first, the device camera captures a frame, which is then processed by the AR software to estimate the camera position regarding to a reference object (e.g., an AR marker). Such an estimation can also make use of internal sensors, which also help to track the reference object. Accurate camera positioning is essential while displaying AR content, since it has to be rotated and scaled according to the scenario. Usually, the image is rendered for the appropriate perspective and it is presented to the user on the display of a device. Note also that interaction with the image is possible through the Interaction Handling module and that, when certain local or remote information is required, the Information Management module is the one responsible for obtaining it.

In this pipeline, the major technological challenges are the internal registration of the objects displayed by the system, the tracking of visual elements and features, and the development of the information display system [21]. In addition, rendering (i.e., photometric registration, comprehensive visualization techniques, or view-management techniques) and real-time data processing are also challenging, since their performance is key when superimposing graphic elements in the environment in a rapid, consistent and realistic way. Thus, accurate, fast and robust registration and tracking are required to develop AR techniques that allow for determining the position and orientation of the observer and the real/virtual objects. Such techniques can be classified into:

- Techniques based on sensors (e.g., Inertial Measuring Unit (IMU), GPS, infrared, ultrasounds).
- Image-based techniques that make use of markers, which can be synthetic or natural.
- Sensor and image fusion techniques.

In addition, there are different techniques for image recognition. Some are based in traditional AR markers, which usually are square or rectangular markers that contain a high-contrast figure similar to a QR code. These type of markers are usually framework-dependent and the software is, in most cases, optimized to recognize them by means of specific algorithms. On the contrary, Natural Feature Tracking (NFT) techniques [22], [23] extract characteristic points from images. Such points are then used to train the AR system to detect them in real time. NFT techniques have the disadvantage of being more computationally intensive and slower than other alternatives, and less effective at long distances, but they provide a seamless integration of the augmented information with the real world, since there is no need for placing an artificial marker visible in the scene [24]. In order to mitigate the disadvantages of NFT, small artificial markers called fiducial markers are often added to the scene to accelerate the initial recognition, improving the performance of the system and reducing the algorithm computational requirements.

## III. IAR FOR INDUSTRIAL APPLICATIONS

The combination of the latest advances in electronics, sensors, networking, and robotics, together with paradigms like IoT, enable the development of advanced applications for industrial systems [25], energy efficiency [26], [27], home automation [28], precision agriculture [29], high-security IoT applications [30]–[34], transportation [35] or for defense and public safety [36], [37].

Among the enabling technologies, IAR has proven to be a suitable tool for the manufacturing strategies proposed by different countries (e.g., Industry 4.0 [38], Industrial Internet of Things [39], Made in China 2025 [40] or smart factories [41]), allowing workers to collaborate among them, to interact with real-time information and to monitor and control systems. Figure 3 summarizes the most relevant industrial tasks and sectors where IAR can bring value.

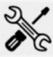

FIGURE 3: Value of IAR across Industry 4.0.

One of the IAR most common applications is the assistance to workers in maintenance/repair/control tasks through instructions with textual, visual, or auditory information [42]. Such an information is rendered ubiquitously, so that the worker perceives the instructions with less effort, thus avoiding the change from a real context to a virtual one where the relevant data is accessed.

Remote assistance is also key when companies have machines installed in remote locations. Such machines need to be monitored, operated and repaired with the minimum amount of people on-site. IAR can help by easing remote collaboration between workers [43], [44]. Augmented communications can also be used for collaborative visualization in engineering processes during stages related to design or manufacturing [45].

Likewise, IAR is helpful for assisting workers in the decision making in real scenarios, combining the physical experience together with the display of information extracted in real time from databases [46]. Furthermore, IAR can provide quick access to documentation like manuals, drawings or 3D models [47]–[49].

Well-trained operators are also essential for productive factories. IAR can help during the training process by giving step-by-step instructions to develop specific tasks. This is especially useful when training workers to operate machinery like the one used for assembly in sequence, what reduces the time and effort dedicated to check manuals [50]. Thus, IAR can reduce the training time for new employees and lower the skill requirements for new hires. In addition, it is possible to adjust the instructions to the experience of the worker, what accelerates the learning process by focusing more on acquiring skills.

IAR training systems are also useful for preserving certain practical knowledge acquired by the most skilled (and usu-ally aged) workers. What happens in different industries is that during the 1960s and 1970s there were massive hirings of workers that are currently retiring [51]. Such workers take with them a lot of experience, knowledge and skills that are difficult to reproduce in traditional ways. Therefore, IAR solutions can include such a knowledge in applications to train new hirings [52], [53].

Another area where IAR can help is marketing and sales, since AR demos are usually really attractive when showing other people a certain scenario or the capabilities of a product. Therefore, AR has the ability to transform the customer experience, enabling the inclusion of different parameters, options, settings or configurations [54]–[57]. Moreover, an improved customer experience is able to reduce the levels of uncertainty about their choices shortening the sales cycle. Furthermore, it can also be used to collect data about product preferences. A good marketing and sales example was created by BMW [58] for a campaign for selling its latest electric vehicles.

After-sales service to the customers can be also enhanced through IAR, since it can guide them through repairs and can connect them with remote experts [59], [60]. Feedback can be also obtained by using an IAR interface, either by receiving it directly from the customers, or by analyzing how such consumers have used the product [61], [62].

The 3D models provided by IAR are a useful tool for engineers while creating and evaluating designs and products [19], [63]. IAR makes it possible to place a virtual object anywhere and observe at full scale whether it fits or not in a specific scenario. Moreover, IAR enables providing on-site CAD model corrections, thus improving accuracy, alignment and other details of the model [64], [65]. In addition, during product the different product manufacturing stages, IAR can help during the quality assurance controls and show performance dashboards [66].

Manufacturing can also be benefited from IAR, which IAR can deliver the right information at the right time to avoid mistakes and increase productivity [67]. This is especially critical in dangerous manufacturing tasks, where a mistake may mean that a worker gets injured or that a costly equipment is damaged. In addition, in such situations an IAR solution could be used as a monitoring and diagnosis tool capturing the information provided by control and management systems, and sensors [68]. For example, Frigo et al. [69] show some use cases in aerospace manufacturing processes.

Assembly is one of the processes whose performance can be improved dramatically by IAR [70]. For example, some researchers have studied different methodologies for developing CAD model-based systems for assembly simulation [71]. In addition, it has already been evaluated the performance of Microsoft HoloLens for assembly processes [72]. Furthermore, a comprehensive review of AR-based assembly processes can be found in [73].

Regarding logistics, IAR can enhance the efficiency of the picking process in warehouse by providing an indoor guidance system [74]. Note that picking items represents from 55 % to 65 % of the total cost of warehousing operations [75], which are still mostly carried out through a pick-by-paper approach. Thus, IAR can give instructions to the workers and guide them using the best route.

Finally, after reviewing the state of the art, the following aspects can be pointed out as essential in order to develop a successful IAR application:

- The use cases and applications selected must provide added-value services.
- Functional discontinuities or gaps in the operating modes that can affect the functionality should be avoided.
- Reduce cognitive discontinuities or differences between old and new work practices. Learning new procedures may hinder the adoption of the technology.
- Reduce physical side-effects caused by the devices on users in the short and long term (e.g., headaches, nausea, or loss of visual acuity).
- Avoid unpredicted effects of the devices on users unfamiliar with the technology, like distractions, surprises or shocks.
- Take into consideration the user perception regarding ergonomic and aesthetic issues.
- Make user interaction as natural and user-friendly as possible, avoiding lapses or inconsistencies.

## IV. IAR FOR SHIPBUILDING

In the last years, several IAR solutions have been presented to help in the accomplishment of daily tasks in a shipyard and in the shipbuilding industry. Some examples are the systems presented in [76]–[78], which facilitate welding processes. In particular, in [76] it is detailed a system that replaces the traditional welder's screen with a helmet that incorporates a display where useful information is projected. Through the screen, a virtual assistant actively suggests corrections during the welding process and indicates possible errors.

Similarly, in [77] it is described a system to control a welding robot inside a ship. The proposed system uses an interface that allows operators to interact with the workspace: the operator receives visual information from a projection system mounted on the robot and controls the robot using a wireless controller. Thus, the system is designed to increase the operator concentration on the actual task, avoiding the distractions related to the use of a traditional screen.

Another relevant welding system is presented in [78], which is aimed at training welders. Such a system consists in a torch, a pair of AR glasses, a motion tracking system and external speakers. Welding is simulated in real time and uses a neural network to determine the quality and shape of the weld based on the speed and orientation of the torch.

Another activity that takes place in a shipyard is spray painting. To simulate the working environment, in [79] it is proposed a system that uses a paint gun as user interface. Such a gun features force feedback and emits painting sound. During training, the student uses this gun to paint on virtual models of steel structures that are shown on the screen of the AR glasses, showing the results immediately, just upon the completion of the exercise.

A system potentially applicable to the shipyard is the one proposed in [80]. It provides AR functionalities on a tablet, allowing the operator to place virtual geo-referenced notes in production modules as a method of communication between workers. The central element of the system adds and processes data from different sources, while the applications running on the tablets retrieve and display such an information for maintenance and for helping in plant processes.

Regarding maintenance in industrial processes of the shipyard, many of them are described in paper or electronic documents that the operator must read and memorize to apply on the ongoing maintenance operations. This task is usually prone to human errors. To avoid them, the system described in [81] presents an IAR-based assistant that makes uses of a tablet to indicate step-by-step instructions that include the necessary information for carrying out an operation. Similarly, in [48] it is detailed an IAR application that provides assistance to military mechanics when performing repair and maintenance tasks on the field, inside an armored vehicle. The prototype proposed uses a wrist control panel and an HMD to augment the mechanic's natural vision with animated text, labels, arrows and sequences to facilitate the understanding, localization and execution of tasks, which include installing and removing locks and warning lights, or connecting wires. Note that, since all those tasks have to be performed in the reduced interior space of an armored vehicle, the IAR prototype makes it easier for the mechanics to locate elements, which is also carried out faster than when making use of traditional documentation systems.

Specific issues arise when building ships for the first time. A common situation is that construction and production processes often overlap over time. When discrepancies between the construction data and the actual ship happen, it is necessary to modify the CAD models. To solve this problem, the AR system detailed in [82] allows the user to visualize the pipe construction data and modify them in the case of misalignment or collision. The modified pipe geometry can be saved and used as an input for bending machines. To ensure the fitting of a pipe, the system integrates an optical measurement tool into the alignment process. All these tasks are performed by using an IAR application in a tablet with a built-in camera that, before the actual installation, allows for visualizing, modifying and verifying pipe segments virtually, being able to interact with them through a touch screen.

Another problem in the naval industry is related to the diagrams used to help managers and engineers to identify the technical requirements and to understand what is going to be built. The correct interpretation of such diagrams is complicated, since they are expressed numerically when they are created during the design stage. In order to overcome this issue, the system detailed in [83] proposes an AR application based on mobile devices that is able to visualize 3D models created from design diagrams. The application does not make use of markers and augments every diagram with a 3D model that is shown through the display of the mobile device. The scanned designs are uploaded to the server with their characteristic points and are stored in databases along with the 3D model. For the recognition of the designs, a picture is taken with the camera of the mobile device and then it is performed a search in the database to locate a match. In the case of finding a match, the 3D model, together with its characteristic points, is transmitted to the mobile device for its interactive visualization.

Finally, it is worth mentioning that companies like Newport News Shipbuilding [84], Index AR Solutions [85] or BAE Systems [86] have already marketed AR solutions focused on specific applications for the naval industry.

## V. IAR USE CASES FOR A SHIPYARD 4.0

After analyzing the state of the art and the processes carried out in workshops and in a ship, the following use cases were selected as the most promising in terms of possible efficiency improvements achieved through the application of IAR:

### 1) Quality control
Its objective is the automation of the quality control processes through the use of computer vision techniques. Product modeling would be carried out by using 3D cameras and reconstruction software. Once the actual model of the product is obtained, it would be possible to detect deviations from the CAD model.

### 2) Assistance in the manufacturing process
This use case is intended to assist operators in the manufacturing process inside a workshop by visualizing 3D models on tangible interfaces located on the workbench. The use of tangible interfaces involves introducing a visual marker/identifier on the working environment that acts as a spatial reference for the IAR system. The marker can be printed on a fixed surface placed on the table of the manufacturing station or on a mobile support that allows for its manipulation by the operator.

### 3) Visualization of the location of products and tools
Thanks to an already developed system based on RFID tags [87]–[89], in Navantia's shipyard in Ferrol it is possible to visualize the 2D-location of products and tools in portable devices such as tablets or IAR glasses (illustrated in Figure 4). This system would make use of sensor values and artificial tags distributed throughout the workshop that allow for determining the positioning of the user in the shipyard.

### 4) Management of warehouses
This use case is intended to assist warehouse operators in the processes of storage, localization, relocation and collection of parts. The immediate advantage of displaying information on IAR devices is a decrease in human errors and in the time associated with the different processes involved in the management of the warehouse.

### 5) Predictive maintenance using data mining
The purpose of this use case is to carry out predictive maintenance actions using data mining techniques. To this end, information related to the quality control of the processes is collected as well as data from sensors in machines and throughout a workshop. Then, the information can be analyzed and shown through IAR devices to the operators.

### 6) Augmented communication
IAR facilitates augmented communication between operators and controllers/experts through portable IAR devices. It enables guidance or resolution of incidents on the spot, sharing the Point of View (POV) of the operator regardless of location, enabling the superposition of information over the actual image seen by the operator, recording annotations and audio/video communications.

### 7) Visualization of installations in hidden areas
This use case consists in the visualization of hidden installations behind bulkheads, roofs or ceilings, by overlapping 3D virtual elements over the real environment. The ideal IAR system would assist operators during the assembly processes, both during the pre-assembly and the block assembly, in order to reveal internal elements that would be difficult to see otherwise. In addition, such a system would ease maintenance or fault repairs. An example of scenario would be one in which, after the detection of a failure in a ship, the operator uses the proposed system to inspect the installations that are invisible to the naked eye in the affected area, deciding in a more agile and fast way the repair operation (Figure 5).

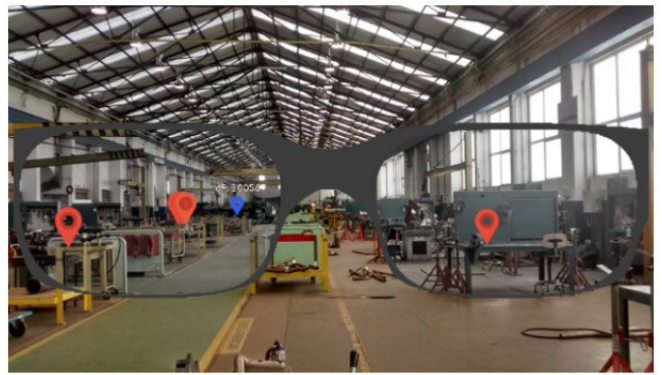

FIGURE 4: Localization of products inside a workshop using an IAR system.

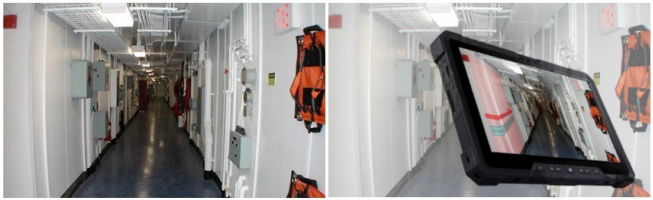

FIGURE 5: a) Real visualization b) Augmented visualization of hidden areas.

#### 8) Remote operation of IIoT and smart connected products/devices

A virtual control panel (digital user interface) can be superimposed on the smart product to be operated directly using an IAR headset, hand gestures, voice commands, or even through another smart product.

## VI. MAIN IAR TECHNOLOGIES FOR SHIPBUILDING APPLICATIONS

### A. HARDWARE DEVICES

Traditionally, assembly and other industrial tasks were constrained due to computational and hardware limitations, thus requiring the use of cumbersome machinery. Tablets and smartphones were an improvement, but they can distract the worker's attention, needing their hands to operate the device when switching between instructions and the industrial process itself. Head-Mounted Display (HMD) devices overcome such distractions by allowing users to view the information in a hands-free manner whilst simultaneously performing an operation. In addition, the glasses can make use of gestures, touch and voice recognition. In fact, the evolution of these latter devices will contribute to increase sales, which are expected to grow from 114,000 units shipped in 2015 to 5.4 million units in 2020 [105].

Another approach would consist in utilizing projectors to display graphical information onto physical objects, known as Spatial Augmented Reality (SAR). SAR detaches the technology from the operators and enables seamless collaboration among them. However, the large extension of a shipyard makes it really difficult and expensive to deploy an effective system that covers most of the production areas, whereas an HMD device can be carried by the operator throughout the shipyard. In addition, recent improvements in HMDs and see-through displays provide appealing features that generate an immersive and personalized user experience that cannot be currently obtained through SAR.

This section is aimed at providing an insight into the current state of IAR technology as far as hardware is concerned. To this end, it has been carried out a comparison of the latest HMD devices, either commercially available or that are scheduled to be launched for sale to the public soon. Table 1 shows a summary of the characteristics of the selected models. Such a table allows for concluding that the general characteristics of these devices are:

- The price of the devices ranges from $ 500 to $ 4,900 (the price of the Daqri helmet is still unknown, but presumably more expensive).
- The average autonomy of these devices in full operation is about 4 hours.
- All devices include 3-axis gyroscopes and IMUs with accelerometers with several DOF (Degrees of Freedom).
- The most commonly embedded sensors measure altitude, humidity, ambient light, pressure, infrared and proximity sensors. In addition, GPS receivers and compasses are also common.
- They include up to 1-2 GB of RAM and up to 128 GB of ROM. Some models support external memory expansion cards. All devices support Wi-Fi and Bluetooth connectivity. Other models, additionally, support LTE.
- All devices except for the FUJITSU Ubiquitousware HMD embed optical see-through displays, whose field of vision, resolution and color depth differ from one model to another.
- The quality of the embedded camera also varies remarkably depending on the model. It should be noted that:
  - The Epson Moverio BT-2000 glasses include a deep-focus camera.
  - The Daqri smart helmet includes an additional thermal camera and an Intel RealSense camera (which is a combination of three cameras: a traditional one and two infrared stereo cameras).
  - Some models have built-in microphone and a headset. Others support such connections via an audio jack.
  - Depending on the model, the navigation can be performed via buttons or gestures, or by using voice commands.

It is worth mentioning that other authors [106], after performing a systematic review, point out five key characteristics that smart glasses should have when implementing applications for smart factories:

- The field of view should be as wide as possible, being 30°(horizontally) the recommended minimum for a good user experience.
- Since smart glasses are meant to be worn during the whole day, they should be as light as possible.
- Batteries should last through the working day.
- The video-based display technologies should be avoided, since they incur in delays that harm the user experience. Optical and retinal projection are then rec-

TABLE 1: Characteristics of the most relevant HMD devices for IAR (first part).

| Product | Weight | Price | Hardware | Sensors | Connectivity | Battery | Display | Input/Output | Accessories | Software |
|---|---|---|---|---|---|---|---|---|---|---|
| ATHEER Air glasses [90] | ~350 g | ~$ 3,950 | NVIDIA Tegra K1 processor, 2 GB RAM Up to 128 GB storage | 9–DOF IMU with accelerometer, gyroscope and magnetometer | Wi-Fi, Bluetooth 4.1, GPS, optional 4G LTE | Built-in 3,100 mAh battery, battery life: ~8 hours | See-through stereo 3D displays with a field of view of 50°(diagonal) | 3D depth, dual 4 MP RGB cameras, directional microphone, touch-pad, gesture control | Replaceable face shields, prescription inserts, extended battery, carrying case, docking station | Android-based AiR OS, collaboration-centric AiR Enterprise Suite. |
| CHIPSIP Sime Glasses [91] | 72 g | ~$ 700 | Newton32 System in Package (SiP), Dual cortex-A9 1.2 GHz, 16 GB DDR3L, 8 GB Flash | 9-DOF IMU and light sensor | Wi-Fi (IEEE 802.11 b/g/n), Bluetooth 4.0, micro-USB | 610 mAh lithium-ion. Battery life: 1 hour | See-through display 800×480. Field of view: 20° (diagonal) | 5 MP@1080p, voice and gesture control | Earphone adapter, USB to micro-USB cable, touchpad | Android 4.4.2 standard w/ SiME SDK |
| DAQRI Smarthelmet [92] | Not found | Not Announced | Intel Core m7-6Y75 processor, 64 GB Solid State Drive, 8 GB RAM, up to 3.1 GHz, Intel HD Graphics 515 | 6-DOF IMU with a 3-axis gyroscope, accelerometer and magnetometer. Barometer and temperature sensors | Wi-Fi (IEEE 802.11ac), Bluetooth 4.2 | Two 5,800 mAh Lithium-Ion batteries | Dedicated vision processor, low-power, robust 6-DOF IMU and positional tracker. Multiple camera array including still, video & depth, Intel RealSense Camera LR200 (depth sensor resolution: 480 × 360, 60 fps; depth range: 0.4-4 m RGB 1080p camera, 30 fps) | Wide-angle AR tracking camera (166 ° diagonal fish-eye Lens, resolution: 640 × 480, 100 fps) thermal camera two USB 3.1 Type C ports, headphone jack | Handheld Bluetooth keyboard/mouse with mini USB, USB-C cable, USB-C Hub-Two Type A, Ethernet, HDMI, USB-C power supply | SDK for application development, C++ API for integration with any rendering engine, unity extension and application templates, DAQRI application suite |
| EPSON Moverio BT-200 [93] | ~88 g Headset and ~124 g Controller | € 699.99 | TI OMAP 4460 1.2 Ghz Dual Core, 1 GB RAM, 8 GB of internal memory, micro-SD (max. 2 GB) / micro-SDHC (max. 32 GB) | GPS in the controller, 3-axis compass, gyroscope and accelerometer in both the headset and the controller | IEEE 802.11 b/g/n with Wi-Fi Miracast, Bluetooth 3.0, USB 2.0 (On-The-Go) | 2720 mAh Li-polymer, battery life: ~6 hours. | Poly-silicon TFT active matrix, 0.42 inch wide panel (16:9), 518,400 dots (960 × 540) 60 Hz. Field of view: 23° (diagonal), Screen size (projected distance): 80 inch at 5 m - 320 inch at 20 m, 24 bit-color (16.77 million colors) | Camera (VGA), microphone, capacitive multi-touch touch-pad, Sound Output with Dolby Digital Plus | 3.5 mm headphone jack, microphone | Android 4.0.4 |
| EPSON Moverio BT-300 [94] | ~69 g (without cable) | € 849 | Intel Atom x5 (1.44 GHz, quad-core), 16-GB internal storage, MicroSDHC (up to 32 GB). | GPS, magnetometer, accelerometer, gyroscope and ambient light sensor. | Wi-Fi (IEEE 802.11a/b/g/n/ac), Bluetooth 4.1, micro-USB | Battery life: 6 hours | Optical see-through, 24-bit HD color display. Field of view: 23 °(diagonal) | Camera: 5 MP, 720 p.4-pin mini-jack | Trackpad | Android 5.1 |
| EPSON Moverio BT-2000 [95] | ~290 g Headset ~265 g Controller without batteries (one battery is approx. 50 g) | €3,146 | TI OMAP 4460 (1.2 GHz, dual-core), 1 GB RAM, 8 GB of internal memory, MicroSD (max. 2GB) / microSDHC (max. 32GB) | GPS, magnetometer, gyroscope and ambient light sensor | LAN wireless, IEEE 802.11 b/g/n with Wi-Fi Direct, Bluetooth 4.0, USB 2.0 (On-The-Go), 4-pin mini jack | 1,240 mAh (×2) Lithium-ion batteries, battery life: ~4 hours | Poly-silicon TFT active matrix LCD Size, 0.42 inch wide panel (16:9) LCD pixel number: 518,400 dots [(960 ×540) ×3], 60 Hz, Field of view: 23 °, 24 bit-color (16.77 million colors), Screen size (projected distance): 64 inch at 4 m | Camera: 5 MP (image and video recording), 0.3 million pixels (depth sensing) | Earphones with microphone compliant with the CTIA standard. IP54 (durable against dust and water) | OS Version: Android 4.0.4. Supported 3D (side-by-side format) |
| FUJITSU Ubiquitousware HMD [96] | Main unit & battery ~305 g, belts ~72 g, helmet attachment clip ~45 g | $ 3,000-3,500 | APQ8026 Quad Core 1.2GHz, 2 GB RAM, 8 GB ROM | 3-axis accelerometer, gyroscope, magnetometer, barometer, ambient light sensor | Wi-Fi (IEEE 802.11a/b/g/n), Bluetooth 3.0 | Battery life:-4 hours | Non-see-through extension positioned in front of one eye 0.4-inch (virtual screen size is 15 inches), 854 × 480 (FWVGA) | Effective pixel count of approximately 8.1 MP | - | Android 4.4 |
| Laster WAVƎ [97] | ~ 100 g | € 1,490 | MiPS M200 1.2 GHz, 8 GB eMMC, 512 Mo LPDDR2 | 3-axis accelerometer, gyroscope and magnetometer | Wi-Fi, Bluetooth 4.0, micro-USB | Battery life: 5 hours, external battery pack | See-through holographic lenses. Field of view: 21 °(diagonal) | Camera: 8 MP, 720 p, 30 FPS. 5 control buttons and touch pad | Integrated speaker, microphone | Android |

TABLE 1: Characteristics of the most relevant HMD devices for IAR (second part).

| Product | Weight | Price | Hardware | Sensors | Connectivity | Battery | Display | Input/Output | Accessories | Software |
|---|---|---|---|---|---|---|---|---|---|---|
| Microsoft HoloLens [98] | 579 g | $3,000 | Custom Microsoft Holographic Processing Unit HPU 1.0, Intel 32-bit architecture, 2 GB RAM, 64 GB storage | IMU, ambient light sensor | Wi-Fi (IEEE 802.11ac), Bluetooth 4.1 LE, Micro-USB 2.0 | Battery life: 2-3 hours with active use, 2 weeks in standby | See-through holographic lenses (waveguides), 2x HD 16:9 light engines, automatic pupillary distance calibration, 2.3 M total light points holographic resolution, 2.5 k light points per radian | 2 MP photos, HD video, external speakers, 3.5mm audio jack, four environment understanding cameras, mixed reality capture, four microphones | Not available | Windows 10 with Windows Store. Human Understanding (spatial sound, gaze tracking, gesture input, voice support) |
| ODG R-7 smartglasses [99] | <170 g | $2,750 | Qualcomm Snapdragon 805 (2.7 GHz quad-core processor), 3 GB LP-DDR3 RAM, 64 GB storage | Integrated IMU with a 3-axis accelerometer, a gyroscope and a magnetometer. Altitude, humidity and ambient light sensors | Bluetooth 4.1 (High Speed, BLE, ANT+), IEEE 802.11 ac, GNSS (GPS/GLONASS) | 1,300 mAH Lithium-Ion, battery life: 1-6 hours | Dual 720p stereoscopic see-through displays at 80 fps, 80% see-through transmission magnetic removable photochromic shields. Field of view: 30 °(diagonal) | Auto-focus camera (1080p@60fps, 720p@120fps), 2 digital microphones (user & environment), magnetic charging port with USB on-the-go, magnetic stereo audio ports with ear buds | Protective case, earbuds, BLE wireless finger controller | ReticleOS™, Android Marshmallow Framework. Developer program & website |
| Optinvent ORA-2 [100] | 90 g | €699 | OMAP 4460 (1.2 GHz, dual-core, 32-bit), 4 GB flash, MicroSDHC (up to 32 GB), 1 GB RAM | GPS, 3-axis accelerometer, gyroscope, magnetometer, ambient light sensor | Wi-Fi (IEEE 802.11b/g/n), Bluetooth, micro-USB | Battery life: 3 hours, external battery pack | Optical see-through, 23-bit full color display, 16:9 aspect ratio. Field of view: 23 °(diagonal) | Camera: 5 MP pictures, 720 p@30 FPS video | Touchpad, ear speaker, microphone. IP65 (durable against dust and water) | Android 4.4.2 |
| Penny C Wear 30 Extended [101] | 65 g tethered computer box, 115 g battery weight | Unknown | Intel Core m5-6Y30, 4 GB, 256 GB | 3-axis gyroscope | Wi-Fi, Bluetooth, micro-USB | Unknown characteristics, allows for using external power banks | Retinal see-through, OLED 854×480 resolution. Field of view: 42° × 25° | Audio and camera not integrated | Jawbone click sensor | Windows and Linux are supported |
| RECON Jet Glasses [102] | 85 g | $499 | 1 GHz dual-core ARM Cortex-A9, 1 GB SDRAM, 8 GB flash memory | 3-axis accelerometer, gyroscope, magnetometer, barometer and infrared sensor | GPS, Bluetooth 4.0, ANT+, Wi-Fi(IEEE 802.11a/b/g/n) and Micro USB 2.0 | Swappable lithium-ion, battery life: ~4 hours | WQVGA 16:9, virtual image appears as 30" HD display at 7'. Power-saving sleep modes including IR-enabled glance detection technology | Optical touch pad, 2-button rocker, Point-of-view photo and video camera | Dual microphones and integrated speaker | Android-based ReconOS 4 |
| SONY SmartEyeglass [103] | ~77 g eyewear and ~45 g controller | €971,61 | Not publicly shared | Accelerometer, gyroscope, magnetometer, ambient light sensor | Wi-Fi (IEEE 802.11b/g), Bluetooth 3.0, micro-USB (only for charging) | Built-in lithium-ion, battery life: +150' of continuous use, ~80' of continuous use when using the camera | Binocular, see-through. Field of view: 20 °(19 °×6 °). Max. brightness: 1,000 cd/$m^2$, Resolution 419×138, monochrome (green) 8-bit | Camera: 3 MP picture, QVGA video without sound, speaker included in the controller, microphone and noise suppression sub-microphone. | Dark filter, camera cover sheet and nose pad | Android 4.4 and above |
| VUZIX M100 glasses [104] | 150 g | €1,079 | OMAP4460 1.2 GHz, 1 GB RAM, 4 GB flash memory, Micro SD (up to 32 GB) | 3-DOF gesture engine, 3-DOF head tracking, 3-axis accelerometer, gyroscope, magnetometer, proximity sensor, ambient light and GPS | Micro USB (for control/power/upgrade), Wi-Fi (IEEE 802.11b/g/n), Bluetooth 4.0 | 550 mAh, battery life: up to 6 hours hands free (display off), 2 hours hands free + display, 1 hour hands free + display + camera + high CPU loading | WQVGA color display 16:9. Field of view: 15 °Brightness: >2000 nits, 24 bit-color | Camera (5 MP stills, 1080p video, 16:9 aspect ratio), ear speaker, dual noise canceling microphone, 4 control buttons. Remote control app (it runs on a paired Android device). Supports customizable voice navigation and gesturing | Mounting options: over head, safety glasses (included), use with left or right eye | Android ICS 4.04, API 15 |

ommended.
- Voice-based interaction is recommended to free the operator hands, but it is currently a challenge to make it work properly in noisy industrial environments.

Furthermore, some additional constraints regarding the implementation of applications for a shipyard environment should be remarked:

- Due to the limited memory of current IAR hardware devices, their use tends to be limited to reduced areas, which have to be mapped (usually on-the-fly) by using different sensors. It is possible to load data dynamically, as the operator enters a new production area, but that complicates the development and requires fast storage and processing hardware.
- Since IAR algorithms make use of computer vision techniques based in thresholding and color measurement metrics, the lighting characteristics (light of the environment, type of light, light temperature) can impact the system performance significantly.
- Most IAR systems are based on detecting physical characteristics of an object or a place. Such a detection is complex and compute intensive and, therefore, powerful hardware is required. Moreover, the detection of physical patterns is also influenced by ambient light, being often even more sensitive to changes in the environment than other marker-based detection techniques.
- Electrical interference produced by the industrial machinery inside the shipyard may affect sensor readings and their accuracy. There are IAR systems that complement visual pattern recognition with embedded sensor measurements. A problem may occur in a shipyard when a system relies on a GPS receiver or in Wi-Fi signals to locate a user or a room, since in certain environments (for instance, indoors or inside a ship under construction) such positioning techniques may not work properly.
- Some IAR systems rely on incremental tracking, combining information collected from different sensors and from the camera. Nonetheless, a dynamic environment like a shipyard workshop or a ship that is being built, where the geometrical structure of the place changes through time, can mislead the IAR system.

Finally, it should be mentioned that the vast majority of current IAR hardware devices can be still considered experimental developments, what makes it difficult their integration with existing IAR frameworks and the implementation of new features. An open-source framework would make it possible to develop modifications to work with different platforms.

### B. SOFTWARE DEVELOPMENT TOOLS FOR IAR

Nowadays, there are many libraries for the development of IAR applications, each implementing certain functionalities required in a specific scenario. In general, for the development of an IAR application it is necessary:

- 2D and 3D graphic libraries, which should enable real-time visualization and overlapping of virtual elements in the field of view.
- Recognition mechanisms to be able to follow objects or to superimpose information on them.
- Speech recognition, which is very useful when the user is not able to interact with physical controls. Gesture recognition is also useful when ambient noise prevents voice recognition.
- Reconstruction of 3D environments to get an understanding of the surroundings.
- Overlapping 3D virtual elements (with or without AR markers).

All these features should be addressed by the different pieces of an IAR Software Development Kit (SDK), which is responsible for interacting with the hardware of the device to obtain data from the environment in order to show contextual information to the user depending on the surrounding scenario.

The most relevant SDKs for developing IAR applications are shown in Table 2. They are compared according to several requirements. First, their license type: open-source (i.e., ARToolkit, ArUco, BazAr, UART, OpenSpace3D), free (i.e., ALVAR, Armedia, Vuforia, Wikitude), or commercial versions (i.e., Armedia, Vuforia, Wikitude, ZappCode Creator). Other criteria are the platforms supported (i.e., Android, iOS, Linux or Windows) and requirements about the marker generation, tracking or the overlaying capabilities.

Beyond the general requirements explained, the SDK should be compatible with the chosen HMD device (i.e., Vuforia is compatible with Epson Moverio, ODG R-7 and HoloLens).

Moreover, the compatibility with Unity should be considered. Unity is currently one of the most advanced game engines in the market and it is possible to use it for developing and deploying IAR applications. Examples of SDKs with this feature are UART, Vuforia, ARKit or ARCore. Additionally, the SDK chosen should consider geo-location support for creating location-based IAR applications, and Simultaneous Localization and Mapping (SLAM) to map an environment and track movements in order to enable indoor navigation. Examples of this feature can be seen in ARKit or Instant Reality. Cloud services and text recognition features should be also taken into account in scenarios where they are required.

At the time of writing this article (December 2017), all of the referred SDKs are available. Nevertheless, the IAR landscape is constantly evolving and many SDKs available in the last years have gone away, changed (e.g., different license type) due to the shifting commercial priorities or have been overridden by newer projects (e.g., ARToolkit has been recently acquired by Daqri).

TABLE 2: Most relevant SDKs for developing IAR applications (first part).

| Product | License type | Platform | Characteristics | Description |
|---|---|---|---|---|
| ALVAR [107] | Free, Commercial SDK | Android, iOS, Windows, Flash | Marker, NFT | Software library for creating virtual and AR applications. Developed by the VTT Technical Research Center of Finland. |
| ARKit [108] | Commercial SDK | Xcode 9 and iOS 11 | Visual Inertial Odometry (VIO), Simultaneous Localization And Mapping (SLAM), Face tracking | Apple's ARKit is in beta version. It uses VIO to accurately track the world around fusing camera sensor data with Core Motion data. These two inputs allow the iPhone X move with a high degree of accuracy without any additional calibration. |
| ARCore [109] | Free, Commercial SDK | Android | Motion tracking, environmental understanding | Google's ARCore is being offered as a developer preview. ARCore apps will run on the Samsung Galaxy S8 and Google Pixel to start, and will come to millions more Android devices running N and later. |
| ARLab [110] | Free, Commercial SDK | Android, iOS | GPS, IMU Sensors, Visual Search | Global reference in the industry. |
| ARmedia [111] | Free, Commercial SDK | Android, iOS, Windows, Flash | Marker, NFT, GPS, IMU | Structured and modular development framework independent of the real-time tracking and rendering engine. Compatible with Epson Moverio BT-200 and BT-2000. |
| ARToolkit [112] | Open Source, Commercial SDK | Android, iOS, Linux OSX, Windows | Marker, NFT | Recently acquired by Daqri. Compatible with Epson Moverio and Daqri Smarthelmet. |
| ArUco [113] | Open Source | Linux, OSX, Windows | Marker | It uses the OpenCV library. |
| Augumenta Interaction Platform [114] | Commercial SDK | Android, Linux, Windows | NFT | It offers gesture control and virtual surfaces. It is compatible with Vuforia. Compatible with ODG R-7, Epson Moverio, Google Glass and ChipSiP SiME. |
| Aurasma [115] | Free, Commercial SDK | Android, iOS | NFT, Visual Search | Owned by HP, it incorporates its own image recognition solution. |
| BazAr [116] | Open Source | Linux, OSX, Windows | NFT | Computer vision library based on feature point detection and matching. It is able to detect and register quickly known planar objects in images. |
| BeyondAR [117] | Free (Apache) | Android | GPS, markers, IMU Sensors | Platform specialized in geolocation-based AR applications on smartphones and tablets. It allows for creating easily 3D objects and for viewing them through a camera. |
| Beyond Reality Face [118] | Comercial SDK | Flash | Face tracking | Its API is small, clear and unified for all platforms. |
| Catchoom [119] | Free, Commercial SDK | Android, iOS | Visual Search | The SDKs licensed by Catchoom Technologies offer tools to connect mobile applications with the CraftAR service in the cloud. |
| IN2AR [120] | Free, Commercial SDK | Flash, Android, iOS | NFT | Cross-platform engine able to detect and estimate the position of images using webcams and mobile cameras. Such an information can be used to superimpose 3D objects or videos on the images to create motion-controlled applications. |
| Instant Reality [121] | Free, Commercial SDK | Android, iOS, Linux, OSX, Windows | Markers, NFT, GPS, IMU sensors, Facial Tracking, Visual Search, ContentAPI, SLAM, TrackerInterface | Developed by Fraunhofer IGD and ZGDV in cooperation with other industrial partners. This framework for mixed reality systems presents interfaces for developers to access components for AR/VR applications. |
| Layar [122] | Free, Commercial SDK | Android, iOS | Natural Feature, GPS, IMU sensors, Visual Search, ContentAPI | Layar allows publishers, advertisers and brands to create interactive AR contents without the need for developing or installing software. |

TABLE 2: Most relevant SDKs for developing IAR applications (second part).

| Product | License type | Platform | Characteristics | Description |
|---|---|---|---|---|
| Mixare (mixed AR Engine) [123] | Open Source (GPLv3) | Android, iOS | GPS | Free AR browser that works as a completely autonomous application and that is also available for developers. |
| OpenSpace3D [124] | Open Source | Multi-platform (Android, iOS, Linux, OSX, Windows) | Markers, Aruco fiducial marker detection | Native support of Google Carboard, HTC Vive, Oculus, Leap motion and other low-cost solutions. It is also available for professional systems like Quad-buffer render and Virtual Reality Peripheral Network (VRPN). |
| SSTT [125] | Proprietary | Android, iOS, Linux, OSX, Windows, Windows Mobile | Markers, NFT | Simplified Spatial Target Tracker (SSTT) is a computer vision based tracking library for AR and mixed reality applications. It uses a WebKit-based browser and adds fast AR NFT tracking to it. |
| UART [126] | Open Source | iOS, OSX, Windows | Markers | Unity AR Toolkit (UART) is a set of plugins for the Unity engine that allows users to develop and deploy AR applications. |
| Vuforia [127] | Free, Commercial SDK | Android, iOS | Markers, NFT, Visual Search | It incorporates computer vision technology to recognize and track planar images and 3D objects in real time. This image registration capability enables developers to position and orient virtual objects in relation to real world images when they are viewed through the camera of a device. Moreover, Vuforia's SDK provides a list of 100,000 commonly used English words that can be incorporated into Text Recognition apps. It is compatible with different Epson Moverio smart glasses, ODG R-7 and Microsoft Hololens. |
| Wikitude [128] | Free, Commercial SDK | Android, IOS, BlackBerry OS | GPS, IMU sensors, ContentAPI | Wikitude is a complete AR solution that includes image recognition, tracking and 3D model rendering. It is compatible with Epson Moverio smart glasses, Sony SmartEyeglass and Vuzix M-100. |
| ZappCode Creator [129] | Comercial SDK | Android, iOS | Markers | Set of content creation tools to develop AR experiences. |

## VII. IAR ARCHITECTURE FOR A SHIPYARD 4.0
### A. TRADITIONAL COMMUNICATIONS ARCHITECTURE

The traditional IAR architecture is composed by the three different layers depicted on the left in Figure 6, which are responsible for data acquisition, data transport, and visualization and interaction with the user.

The Visualization and Interaction System (VIS) layer is composed by HMD, HHD and spatial display devices, as well as by human-interaction interfaces through which the operator, located in the ship under construction, the workshop or the shipyard, is able to interact with the system.

The Data Transport System (DTS) is responsible for collecting the information obtained by the Data Acquisition System (DAS) and transmitting it from the cloud infrastructure to the location where operators are using the IAR system. This process involves certain difficulties due to the environmental conditions: the shipyard's structural barriers and the large number of metal parts that impact wireless communications performance. Moreover, it has to be taken into account the electrical interference produced by the industrial machinery used for the different production processes.

It is worth mentioning that the communications between each IAR device and the cloud layer can be performed through Wi-Fi connections considering that many shipyards have already deployed an IEEE 802.11 b/g/n/ac infrastructure. However, note that communications inside a ship suppose a challenge for electro-magnetic propagation due to the presence of numerous large metal elements, so other technologies would need to be further studied in such environments. Considering that electrical wiring is usually deployed for temporary lighting during the ship construction, Power Line Communication (PLC) technology can be used to transmit information, although it should be noted that the network speed could be influenced by the electrical interference coming from electric circular saws and other tools that demand high-current peaks. Thus, a PLC system would consist of two modules connected electrically to the same power phase. The first module would have direct connection to the DAS via Ethernet. The second PLC module would be in the area of limited connectivity and would be coupled with a high-speed Wi-Fi access point to which the VIS can be connected to.

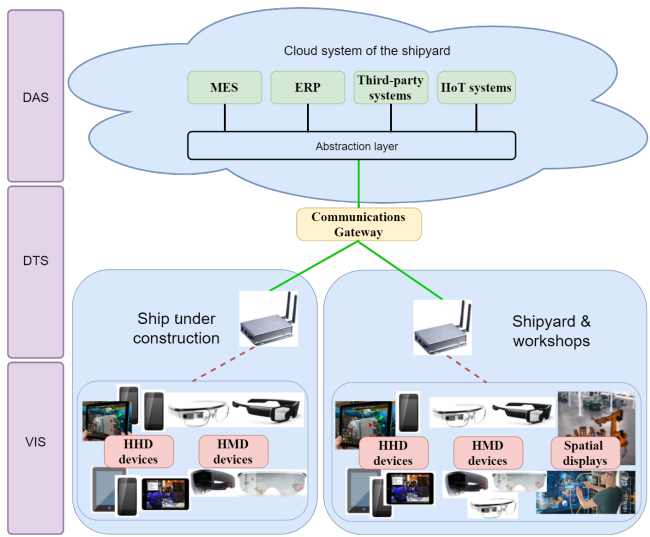

FIGURE 6: Traditional IAR architecture.

Regarding the DAS layer, which is hosted in the shipyard's cloud infrastructure, it can obtain from the Manufacturing Execution System (MES) data about work orders and the elements to be installed.

Note that the described IAR architecture, although it has been previously implemented by making use of a central server, a cloud or a PC-cluster [43], [130], [131], it has certain limitations when applied to IAR systems. Since IAR devices project information dynamically on a display, they require it to be loaded and rendered as fast as possible to provide a good user experience. For this reason, the content to be displayed is usually stored locally, either in static or dynamic memory, which guarantees in most cases low-latency responses when loading data. However, although the local-storage approximation works fine in certain scenarios that only need static content, the information that flows through smart factories and other Industry 4.0 environments is dynamic, especially in cases when showing data that is collected from IIoT networks or other third-party systems (e.g., from the Enterprise Resource Planning (ERP) or the Product Lifecycle Management (PLM)).

In addition, this traditional architecture can be enhanced by adding a proxy server that caches dynamically IAR information. Thus, when the IAR solution predicts that certain content will be required soon, a parallel process of the IAR device can download the data in the background. Nonetheless, such an approximation has several drawbacks. First, IAR mobile devices are usually limited in terms of computational power and battery life, so multi-thread or multi-process applications may reduce the overall performance and derive into increasing energy consumption. Second, caching algorithms involve certain degree of complexity, since content has to be synchronized periodically to guarantee that the latest version is the one loaded. And third, the prediction algorithm has to be accurate enough because unnecessary file exchanges imply wasting resources and energy.

### B. ADVANCED COMMUNICATIONS ARCHITECTURES

Due to the constraints introduced by cloud-centered architectures in IAR applications, several researchers have studied alternatives that improve latency response and rendering performance. The most common approximation involves the use of local high-end PCs that store content and process video information [132], [133].

In the last years it has been proposed the use of the Edge Computing paradigm to enhance IAR systems [17], [45]. Such a paradigm can be implemented through different approximations, including basically Fog Computing, Cloudlets and Mobile Edge Computing [134]. All these variants have the same objective: to move the cloud processing power to the edge of the network, as close as possible to the IAR devices. In the case of IAR Fog Computing based solutions, they make use of nodes scattered throughout a factory or industrial environment that process and store the information received from IAR devices before forwarding it to a Cloud. In the same way, the data received from the Cloud can be cached locally and then delivered to IAR devices on demand. The Mobile Edge Computing paradigm is similar, but the nodes are deployed in cellular network base stations. Regarding Cloudlets, they use similar hardware to Cloud Computing servers, but at a low scale and close to the user IAR devices.

There are not many examples of IAR Edge Computing based systems, being currently the most relevant approach the one presented in [45]. In such a paper it is proposed an architecture that offloads the demanding real-time IAR algorithms to a high-end PC, which enables for decreasing end-to-end latency for video transmissions. Nonetheless, the authors conclude that their system works fine for handheld IAR devices, but that higher bit-rate communication interfaces and smaller network latencies are required for most HMD solutions. Therefore, the use of IAR Edge Computing still has to be further studied.

### C. PROPOSED ARCHITECTURE FOR FUTURE SHIPBUILDING IAR APPLICATIONS

Taking into account the considerations mentioned in the previous subsections, it can be proposed an IAR architecture that both reduces latency response and on-line rendering time. The architecture, depicted in Figure 7, makes use of Fog Computing nodes and Cloudlets. Fog computing nodes are ideal for providing IAR services thanks to its ability to support physically distributed, low-latency and QoS-aware applications that decrease the network traffic and the computational load of traditional Cloud computing systems. Cloudlets offload the rendering from resource-constraint IAR devices and also reduce latency response with respect to a Cloud. Such an Edge Computing architecture is composed by three layers.

The layer at the bottom is the Node Layer, which includes all the IAR devices that interact with the services provided by the Edge Computing Layer. The Edge Computing Layer also exchanges data with the shipyard's IIoT ecosystem, which is composed by all the elements that can be monitored and/or controlled remotely, like sensors, actuators, robots or additive manufacturing devices.

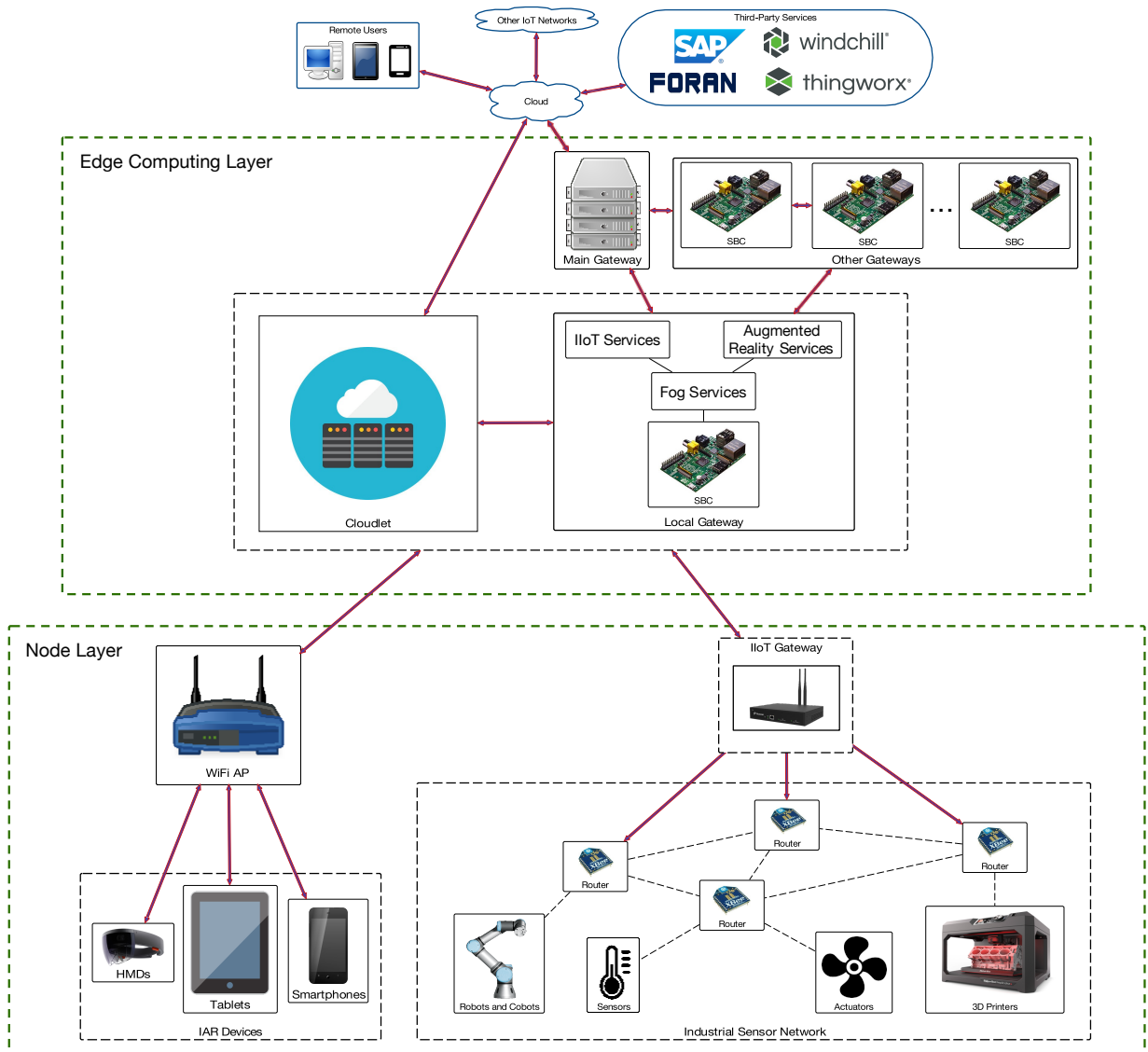

FIGURE 7: Architecture of the IAR system proposed.

It is worth emphasizing that the architecture connects IAR devices with the Edge Computing Layer through a WiFi AP because it is already deployed in many industries and thus it is highly likely that an IEEE 802.11 standard (i.e., IEEE 802.11 a/b/g/n/ac/ad) would be used. However, any other wireless communication technology could be used in the architecture and further research should be performed about this topic in shipyards, especially inside ships that are being built, since electro-magnetic propagation is a challenge in such environments with so many metal elements. One alternative that is currently being evaluated by Navantia is PLC, but its application still has to be studied in detail in situations where electric interference from saws and other tools occur.

The Edge Computing Layer is in the middle of the architecture, and it is actually divided into two sub-layers that can interact with each other, but whose objective is different. The Fog Computing sub-layer is composed by one or several Single-Board Computers (SBCs) that are installed in fixed positions throughout the shipyard workshops and in a ship. Each SBC acts as a gateway and provides fog services. In the case of the IAR service, it supplies IAR devices with localized data and responds faster than the cloud, thus acting as a proxy caching server for IAR data. The second sub-layer is composed by Cloudlets, which are local high-end computers specialized in rendering and performing compute intensive tasks. Note that shipbuilding requires handling sophisticated 3D CAD models that may be too complex for resource-constrained IAR devices, so, instead of rendering the objects locally in every IAR device, the tasks are delegated to a cloudlet that then delivers the resulting image to the IAR device faster than the cloud. In addition, cloudlets can also perform other processing tasks that are too heavy to be performed in an SBC and that require a fast response.

The cloud is at the top of the architecture. It receives, processes and stores data from the Edge Computing Layer. The cloud also provides third-party services to IAR devices. For instance, in the case of a shipbuilder like Navantia, such services include the access to the content of the ERP (SAP ERP), the CAD models (through FORAN), the information of the PLM (Windchill) and to the IIoT platform (ThingWorx).

Finally, note that a modern IAR architectures have also to take security into account. This aspect is often neglected, but the future dependency on IAR systems make security essential. This has been addressed recently by some researchers [17], but further study is required, especially in industrial environments.

## VIII. CONCLUSIONS

This article reviewed the different aspects that influence the design of an IAR system for the Industry 4.0 shipyard, considering diverse scenarios like workshops and a ship. It was first presented a brief overview of the underlying technologies and the main industrial IAR applications. Next, the article analyzed the most relevant academic and commercial IAR developments for the shipbuilding industry. Then, different use cases of IAR for a shipyard were described and a thorough review of the main IAR hardware and software solutions was presented. After such a review, it can be concluded that there are many options for developing software IAR developments, but IAR hardware, although it has progressed a great deal in the last years, it is still not ready for a massive deployment like the one required by the Industry 4.0 shipyard.

Regarding the communications architecture for an IAR system, it was analyzed the traditional version and it was proposed an enhanced three-layer Edge Computing architecture based on Cloudlets and on the Fog Computing paradigm, which allows for supporting physically distributed, low-latency and QoS-aware applications that decrease the network traffic and the computational load of traditional cloud computing systems. Moreover, considering that fog gateways are usually constraint in terms of computing power, if an IAR system demands real-time rendering or compute-intensive services, Cloudlets are necessary.

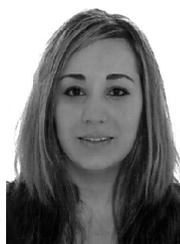

PAULA FRAGA-LAMAS (M'17) received the M.Sc. degree in Computer Science in 2008 from University of A Coruña (UDC) and the M.Sc. and Ph.D. degrees in the joint program Mobile Network Information and Communication Technologies from five Spanish universities: University of the Basque Country, University of Cantabria, University of Zaragoza, University of Oviedo and University of A Coruña, in 2011 and 2017, respectively. Since 2009, she has been working with the Group of Electronic Technology and Communications (GTEC) in the Department of Computer Engineering (UDC). She is co-author of more than thirty peer-reviewed indexed journals, international conferences and book chapters. Her current research interests include wireless communications in mission-critical scenarios, Industry 4.0, Internet of Things (IoT), Augmented Reality (AR), RFID and Cyber-Physical systems (CPS). She has also been participating in more than twenty research projects funded by the regional and national government as well as well as R&D contracts with private companies.

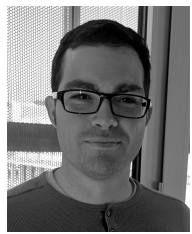

TIAGO M. FERNÁNDEZ-CARAMÉS (S'08-M'12-SM'15) received his MSc degree and PhD degrees in Computer Science in 2005 and 2011 from University of A Coruña, Spain. Since 2005 he has been working with the Department of Computer Engineering at the University of A Coruña. His current research interests include IoT systems, Industry 4.0, Augmented Reality, RFID, wireless sensor networks, embedded systems and wireless communications.

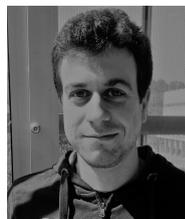

ÓSCAR BLANCO-NOVOA received his B.Sc. in Computer Science in 2016 with mention in computer engineering and Information Technology at the University of A Coruña (UDC). During the last years in college combined his studies with a job as a software engineer at a private company. Currently he is studying his Master's degree in Computer Science and works at the Group of Electronic Technology and Communications (GTEC) in the Department of Computer Engineering (UDC). His current research interests include Energy Control Smart Systems, Augmented Reality and Industry 4.0.

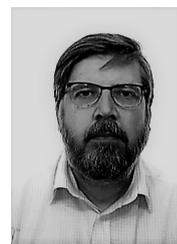

MIGUEL A. VILAR-MONTESINOS received the M.Sc. degree in Electrical Engineering in 1992 from University of Vigo. He started his professional career as an electrical engineer at Astano, subsequently leading to Information Technologies. For the last twelve years he was responsible for Engineering Systems at Navantia and he is currently head of the department of Digitalization Projects. Due to his position, he collaborates with University of A Coruña research on Industry 4.0, Augmented Reality, Internet of Things and RFID systems.

∙ ∙ ∙